\documentclass[11pt]{article}
\usepackage[margin=1in]{geometry}
\usepackage{fix-cm}
\RequirePackage{amsthm,amsmath,amsfonts,amssymb}
\RequirePackage[authoryear]{natbib}
\usepackage[dvipsnames]{xcolor}
\RequirePackage[colorlinks,citecolor=blue,urlcolor=blue]{hyperref}
\RequirePackage{graphicx}
\newcommand{\jianrevise}[1]{\textcolor{black}{#1}}
\newcommand{\mrevise}[1]{\textcolor{black}{#1}}
\usepackage{enumerate}
\usepackage{hyperref}
\usepackage{tabularx}
\usepackage{rotating}
\usepackage{newtxtext}
\usepackage{natbib}

\theoremstyle{plain}

\theoremstyle{remark}

\begin{document}

\title{Statistical Opportunities in Neuroimaging}

\author{
Jian Kang$^{1}$
\and
Thomas Nichols$^{2}$
\and
Lexin Li$^{3}$
\and
Martin A. Lindquist$^{4}$
\and
Hongtu Zhu$^{5}$\thanks{Corresponding author: htzhu@email.unc.edu}\\[5mm]
$^{1}$Department of Biostatistics, University of Michigan, Ann Arbor\\
$^{2}$Big Data Institute, University of Oxford\\
$^{3}$Department of Biostatistics, University of California, Berkeley\\
$^{4}$Department of Biostatistics, Johns Hopkins University\\
$^{5}$Department of Biostatistics and Biomedical Imaging Research Center,\\
University of North Carolina at Chapel Hill
}

\date{}

\maketitle


\begin{abstract}
Neuroimaging has profoundly enhanced our understanding of the human brain by \mrevise{characterizing} its structure, function, and connectivity through modalities like MRI, fMRI, EEG, and PET. These technologies have enabled major breakthroughs across the lifespan, from early brain development to neurodegenerative and neuropsychiatric disorders. 
Despite these advances, the brain is a complex, multiscale system, and neuroimaging measurements are correspondingly high-dimensional. This creates major statistical challenges, including \mrevise{ measurement} noise, motion-related artifacts, substantial inter-subject and site/scanner variability, and the sheer scale of modern studies.
 This paper explores statistical opportunities and challenges in neuroimaging across four key areas: (i) brain development from birth to age 20, (ii) the adult and aging brain, (iii) neurodegeneration and neuropsychiatric disorders, and (iv) brain encoding and decoding. After a quick tutorial on   major imaging technologies, we review cutting-edge studies, underscore data and modeling challenges, and highlight research opportunities for statisticians. We conclude by emphasizing that close collaboration among statisticians, neuroscientists, and clinicians is essential for translating neuroimaging advances into improved diagnostics, deeper mechanistic insight, and more personalized treatments.
\end{abstract}



\section{Introduction}

\jianrevise{This review focuses on commonly used imaging modalities in human brain research and outlines the statistical challenges and opportunities arising from these data.}
Neuroimaging has fundamentally transformed our understanding of the human brain, offering \mrevise{detailed insights}  of its structure, function, and \mrevise{connectivity}. Essential tools like magnetic resonance imaging (MRI), functional MRI (fMRI), electroencephalography (EEG), and positron emission tomography (PET) have become cornerstones in neuroscience \citep{Filippi2015}. These methods allow for 
\jianrevise{time-varying measurements of brain structure, function, and connectivity}, 
leading to groundbreaking discoveries in brain development, neurodegenerative and neuropsychiatric disorders, and cognitive function \citep{Gibby2015}.

\jianrevise{Neuroimaging data are inherently complex, reflecting both the brain’s underlying biology and the measurement process, which depends on imaging hardware, acquisition parameters, and multiple sources of noise such as thermal fluctuations and subject motion.}
Brain structure and function also vary substantially across individuals due to factors such as age, genetics, ethnicity, disease status, and external influences including lifestyle and environmental exposures.  However, \mrevise{variability introduced by} imaging devices, acquisition settings, and measurement noise can \mrevise{mask} the subtle brain changes  of  scientific interest, making it difficult to disentangle the effects of key predictors. 
 Figure \ref{Figure1IPA} outlines four major areas of individual-level image processing analysis (IPA): image reconstruction, image enhancement, image registration and image segmentation. These topics are central to improving the quality and interpretability of neuroimaging data. Effectively processing individual neuroimaging data  itself presents significant challenges and opportunities for statisticians.  
For  comprehensive reviews of these IPA techniques, see \cite{zhou2021review, shen2017deep,zhu2023statistical}.

Over the past decade, large-scale neuroimaging studies, such as the UK Biobank (UKB), have collected neuroimaging data alongside genetic information and electronic health records, offering valuable insights into brain development and brain-related disorders \citep{miller2016multimodal, littlejohns2020uk}. However, the complexity and scale of these datasets present significant challenges~\citep{lindquist2025statistics}. 
\jianrevise{Traditional statistical models often struggle in this high-dimensional setting for several reasons. The unprecedented scale and dimensionality of modern neuroimaging data impose substantial computational burdens, while the brain’s inherent nature as a complex object, which is characterized by intricate, non-linear spatial and temporal dependence structures, frequently falls outside the scope of classical frameworks. These frameworks often fail to account for the brain's multi-scale organization, multimodal measurements, and the nonstandard covariates inherent in contemporary studies. 
Even when such models remain statistically valid, high noise levels and limited effective sample sizes can result in unstable or weakly informative estimates of these sophisticated neural patterns. These persistent challenges necessitate the development of robust, scalable methods specifically engineered to handle the unique architectural and statistical nuances of the brain as a dynamic and multifaceted biological system.}
Integrating neuroimaging with other data requires innovative and sophisticated analytical approaches \citep{zhu2023statistical, ombao2016handbook,marron2021object}. These methods must effectively bridge gaps between diverse data types while accounting for the inherent variability in each. Looking ahead, the next decade is poised to bring the development of novel statistical learning methods tailored to these complexities, driving advancements in neuroimaging data analysis (NDA) and deepening our understanding of the brain and its role in brain-related disorders.

\begin{figure}
\includegraphics[width=0.95\textwidth]{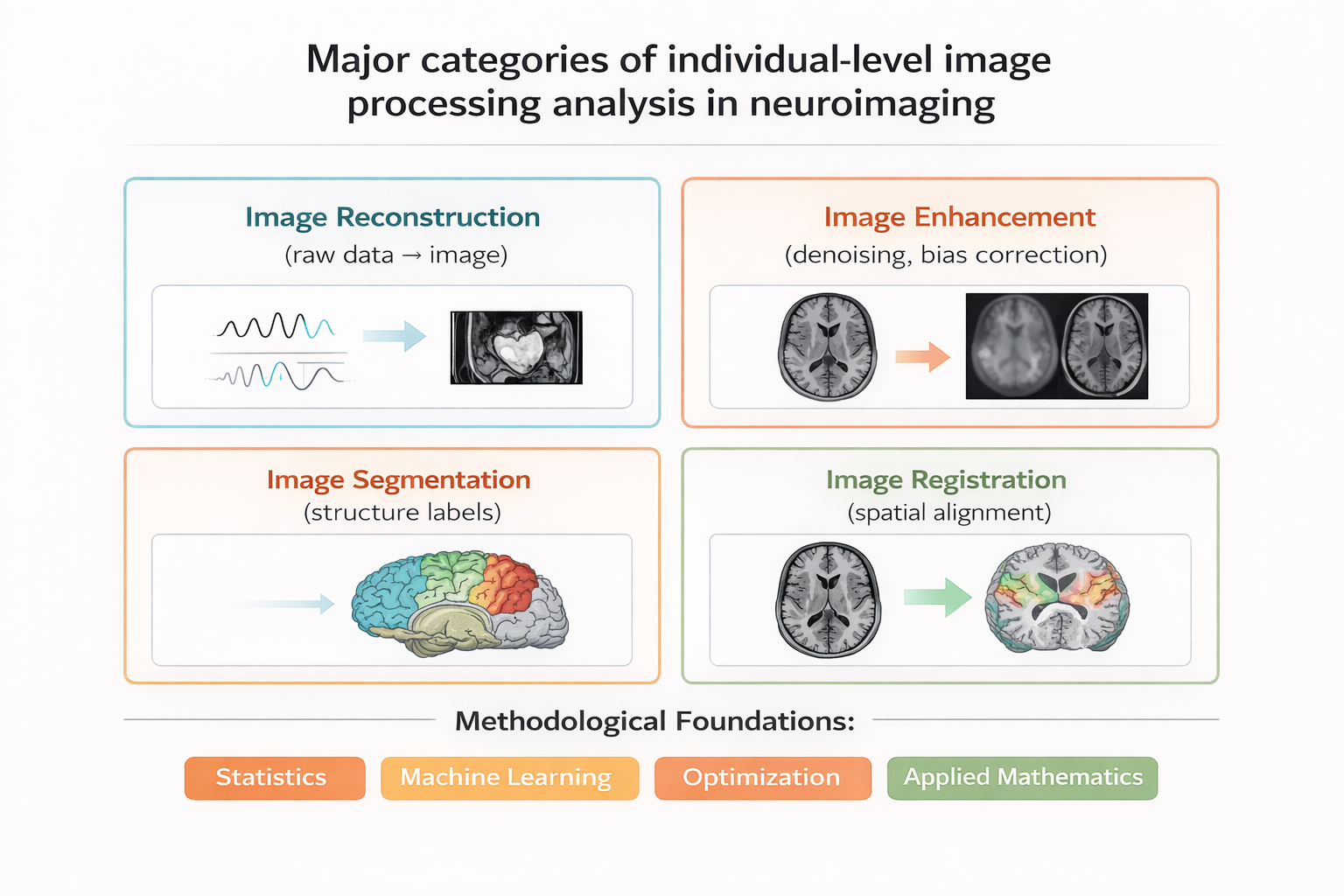}
\caption{\jianrevise{The four core MRI analysis components: image reconstruction (forming images from raw data), image enhancement (noise and artifact reduction), image segmentation (delineating structures of interest), and image registration (spatial alignment across subjects, time points, or modalities). These categories represent a conceptual taxonomy rather than an ordered processing pipeline. The bottom row indicates methodological foundations, including statistics, machine learning, optimization, and applied mathematics, that support all four components.
}}
\end{figure} \label{Figure1IPA}


\jianrevise{This paper examines statistical opportunities and challenges in neuroimaging through four substantive case studies: brain development from birth to age 20, the adult and aging brain, neurodegenerative and neuropsychiatric diseases, and brain encoding and decoding. Section 2 provides foundational background on major neuroimaging modalities and image-processing pipelines, establishing the shared measurement and data-analytic context for the remainder of the paper. Sections 3–6 then use four case studies to illustrate how core statistical challenges, such as high dimensionality, subject heterogeneity, longitudinal missingness, and multimodal data integration, manifest across diverse scientific settings, while also highlighting issues unique to each case study. 
By distinguishing cross-cutting from case-specific challenges, we identify methodological opportunities for statisticians to advance neuroimaging research, improve diagnostic tools, deepen understanding of brain mechanisms, and support the development of personalized treatments. Progress in this interdisciplinary area also requires close collaboration among statisticians, neuroscientists, clinicians, and imaging scientists, including physicists, biomedical engineers, and computer scientists, to fully realize the potential of neuroimaging data.}

\section{Imaging the Human Brain}
Grasping the brain's role in behavior, cognition, \jianrevise{emotion} and personality is essential for understanding neurodegenerative and neuropsychiatric disorders. \jianrevise{Statisticians, working alongside researchers such as physicists, psychologists, and neurologists, can more effectively contribute to interdisciplinary research by understanding basic brain structure and function, as well as the imaging techniques used to study these processes.}

\subsection{Brain basics}

\jianrevise{The brain is organized into three primary compartments: gray matter (GM), white matter (WM), and cerebrospinal fluid (CSF). GM is densely packed with neuronal cell bodies, dendrites, and synapses, forming the cerebral cortex and subcortical nuclei. In contrast, WM consists of myelinated axons and oligodendrocytes that serve as the structural 'wiring' between neural regions. Morphological changes in these compartments are key biomarkers for health and pathology; for instance, healthy aging is characterized by progressive cortical thinning and ex vacuo ventricular expansion, while neuroinflammatory conditions like multiple sclerosis are marked by white matter demyelination and focal axonal lesions.}

\subsection{\jianrevise{Whole-brain measurements
and models}}

\jianrevise{Neuroscience has advanced substantially through a diverse array of neuroimaging modalities, each engineered to probe specific facets of brain architecture and physiology. Electrophysiological methods, such as electroencephalography (EEG), electrocorticography (ECoG), and magnetoencephalography (MEG), capture neuronal currents via electrical potentials or magnetic fields, offering exceptional temporal resolution. In contrast, nuclear medicine techniques, including positron emission tomography (PET) and single-photon emission computed tomography (SPECT), utilize radiotracers to characterize metabolic and molecular processes.}

\jianrevise{
Magnetic resonance-based techniques, spanning structural MRI (sMRI), diffusion MRI (dMRI), and diffusion-weighted imaging (DWI), exploit the intrinsic magnetic properties of tissue to yield high-resolution anatomical and microstructural maps. Hemodynamic-based methods, such as functional MRI (fMRI) and functional near-infrared spectroscopy (fNIRS), indirectly infer neural activity by monitoring fluctuations in blood oxygenation. Finally, computed tomography (CT) leverages X-ray attenuation to quantify tissue density, serving as a critical clinical tool for detecting acute structural abnormalities like hemorrhage. While these macroscale techniques provide invaluable insights, this discussion excludes cellular- and circuit-level methods such as calcium imaging or microscopy \citep{Phelps2000MolecularImaging, Camargo2001BrainSPECT, Zhang2012NODDI}.
}

\jianrevise{Building on these imaging techniques, recent advancements have led to the development of sophisticated whole-brain models that provide a comprehensive view of brain structure and function~\citep{Biswal1995RestingFC,Buckner2008DMN,lindquist2025statistics}. These models typically consist of three key elements:
\begin{enumerate}[(i)]
  \item \textbf{Brain parcellation:} Brain regions are delineated based on structural and/or functional markers, providing a framework for modeling large-scale brain organization~\citep{Buckner2008DMN,Wardlaw2013STRIVE}.
  \item \textbf{Anatomical connectivity matrix:} This matrix encodes the neural pathways linking brain regions and thereby characterizes structural connectivity across the brain~\citep{Zhang2012NODDI,Mueller2005ADNI}.
  \item \textbf{Local dynamics and interaction terms:} These components describe regional activity and interactions among brain regions, supporting the study of large-scale brain dynamics and interactions relevant to cognition~\citep{Biswal1995RestingFC,Buckner2008DMN}.
\end{enumerate}
By integrating these components, whole-brain models can support the analysis and prediction of large-scale brain activity patterns \citep{Shafto2014CamCANProtocol,Taylor2017CamCANRepository}, enhancing our ability to study brain organization, dysfunction, and response to intervention. These models also support advances in neuromodulation and other therapeutic strategies~\citep{Barker1985TMS,OReardon2007rTMSDepression}.}


\begin{figure}
\begin{tabular}{cc}
\includegraphics[height=4cm]{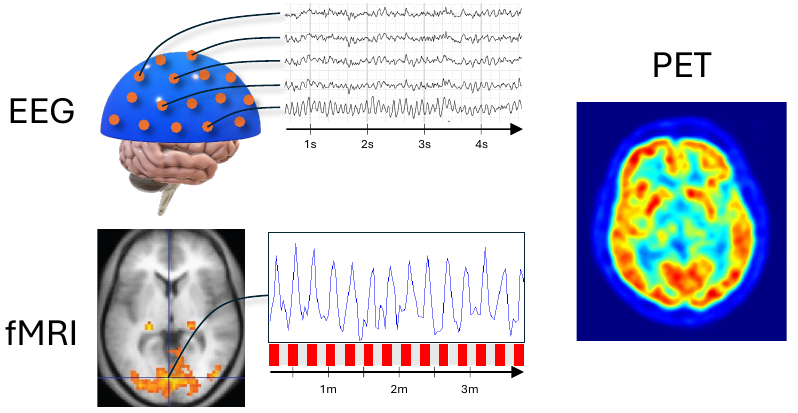}
&
\includegraphics[height=3cm]{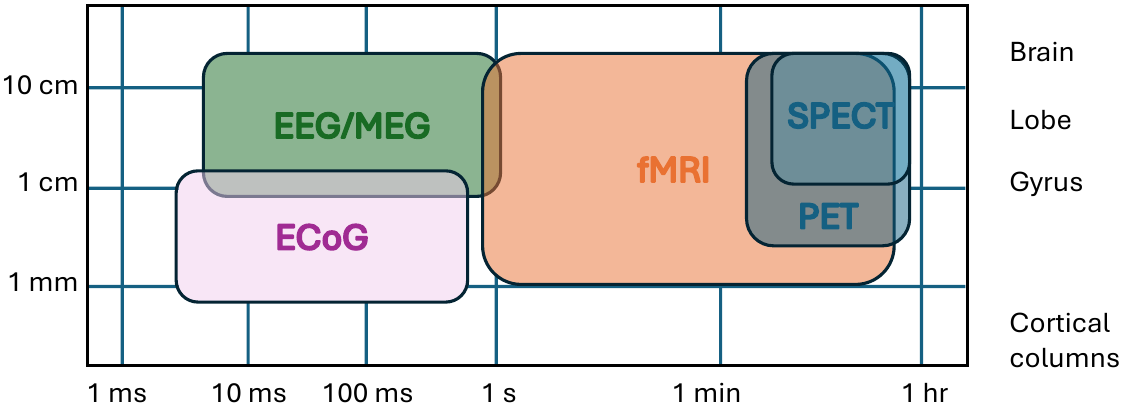}
\end{tabular}
\caption{Left: Illustration of EEG, fMRI and PET; while both EEG and fMRI capture dynamic brain function, EEG has millisecond resolution while fMRI is 1-2s resolution; PET captures a single time point of brain function but can image a different physiological processes depending on the tracer used. Right: Illustration of spatial and temporal resolutions for different imaging modalities. (EEG traces by Andrii Cherninskyi, CC BY-SA 4.0, https://commons.wikimedia.org/w/index.php?curid=44035074; resolution figure after \cite{Sejnowski2014}.)}
\end{figure}

\subsection{Electrophysiology Methods}

Electrophysiology encompasses various methods used in both humans and animals, with EEG and MEG being the most commonly used. EEG measures voltage potentials on the scalp generated by neuronal activity, while MEG detects magnetic fields produced by these neuronal currents. EEG systems may use as few as 20 or as many as 256 electrodes, whereas MEG systems, requiring super-cooled magnetometers, are more cumbersome and expensive. Both methods offer excellent temporal resolution in milliseconds but are limited in spatial resolution due to the number of sensors and head conductivity properties.
\jianrevise{Additionally,  ECoG represents an invasive approach typically used during epilepsy surgery, in which electrodes are placed directly on the cortical surface. By avoiding signal distortion due to the skull and scalp, ECoG offers both excellent temporal resolution and improved spatial resolution relative to noninvasive electrophysiological methods. However, in contrast to whole-brain imaging modalities such as MRI, its spatial coverage is limited to regions where electrodes can be safely implanted, which constrains its field of view.}


\subsection{Nuclear Medicine Imaging}

PET and SPECT are crucial neuroimaging techniques that provide insights into brain function and pathology. PET, one of the earliest methods for creating tomographic brain images, uses radiotracers like fluorodeoxyglucose (FDG) to map glucose metabolism, aiding in the diagnosis and monitoring of neurodegenerative diseases~\citep{Phelps2000MolecularImaging}. It produces images by detecting gamma rays emitted from positron-electron annihilations. PET also enables detection of beta-amyloid plaques with tracers like Pittsburgh Compound B, florbetapir, and florbetaben, which are crucial for Alzheimer's research~\citep{Klunk2004PIB,Clark2011FlorbetapirAutopsy}. SPECT uses a collimator to focus radiation and, while offering lower resolution,  is valuable for blood flow imaging with tracers such as Technetium-99m. Despite using radioactivity, both methods are indispensable for studying specific metabolic pathways and assessing neurological conditions~\citep{Camargo2001BrainSPECT}. \jianrevise{Together, PET and SPECT provide complementary tools for studying brain metabolism, molecular pathology, and perfusion-related dysfunction, making them valuable for investigating neurological conditions characterized by metabolic or vascular alterations.}

\subsection{Magnetic Resonance Imaging (MRI)}

MRI produces detailed 3D images of the brain using a grid of voxels, which can range from 1 to 4 mm in size. Although these images are 3D, the data are often acquired in 2D slices, with dimensions varying from 64$\times$64 to 256$\times$256 pixels, and the total number of slices can range from 50 to over 100.
MRI is incredibly versatile and generates different types of brain images using the same hardware.
{Structural MRI} provides clear images that differentiate between GM, WM, and cerebral structures like ventricles. \jianrevise{The most common contrasts are T1-weighted (T1w) images, which emphasize differences in longitudinal relaxation time and provide high gray–white matter contrast, and T2-weighted (T2w) images, which emphasize transverse relaxation time and are particularly sensitive to fluid and pathology.}
{Functional MRI} (fMRI) tracks brain activity by detecting changes in the blood oxygen level dependent (BOLD) signal. It is used to analyze brain responses during specific tasks or to observe the brain at rest, focusing on changes over time and correlations between different brain regions, respectively.
Diffusion MRI (dMRI) measures the diffusion patterns of water molecules in brain tissue, which helps infer the integrity of white matter and map neural tracts. This technique employs models like the Gaussian diffusion model to assess the directionality of water diffusion, although more sophisticated models, \jianrevise{such as neurite orientation dispersion and density imaging (NODDI), are also commonly used \citep{Zhang2012NODDI}}. \jianrevise{From a data acquisition perspective, MRI scanners do not sample images directly in the spatial domain. Instead, signals are acquired sequentially in k-space, the frequency domain representation of the image, and the final image is obtained via inverse Fourier transformation. As a result, temporally localized disturbances, such as brief head motion or hardware-related noise, can manifest as sparse corruption in k-space, which may translate into spatially widespread or structured artifacts in the reconstructed image. This acquisition–reconstruction interplay has important implications for noise modeling and artifact correction in downstream statistical analyses.}
These MRI techniques collectively offer comprehensive insights into both the anatomical structure and functional dynamics of the brain, making them indispensable in both clinical and research settings.

\

\subsection{functional Near-Infrared Spectroscopy (fNIRS)}

\jianrevise{Brain function can be imaged using functional near-infrared spectroscopy (fNIRS), which exploits the oxygenation-dependent optical properties of blood. By using near-infrared light, fNIRS distinguishes between oxygenated and deoxygenated hemoglobin and measures the vascular hemodynamic responses associated with neural activity~\citep{Ferrari2012fNIRS}. In this respect, fNIRS captures the same underlying hemodynamic signal as fMRI, but relies on optical rather than magnetic properties. Because near-infrared light can penetrate the scalp and skull, fNIRS is well suited for monitoring activity near the cortical surface. However, due to strong optical scattering and absorption in biological tissue, fNIRS has very limited penetration depth and provides little to no sensitivity to deeper brain structures, making it unsuitable for imaging subcortical regions. While simple fNIRS systems may use only a small number of channels, more advanced systems can include 128 or more channels, though their spatial resolution remains substantially lower than that of fMRI.
}

\subsection{Computed Tomography (CT)}

CT uses a rotating source and array of detectors to measure the attenuation of x-rays through the body in order to reconstruct 3D images.  While CT is a workhorse of clinical care,  it is generally inferior to MRI for brain imaging because there is little contrast in the  attenuation coefficients of different types of healthy brain tissue. \jianrevise{However, CT can be used to image stroke lesions, and recent work has advocated leveraging the vast corpus of routinely acquired clinical CT scans for opportunistic screening and secondary analyses \citep{Pickhardt2022OpportunisticCT}}.

\subsection{Image Processing Analysis}

 Imaging data, such as MRI, require extensive image preprocessing and analysis before they can be meaningfully compared across subjects. The primary goal of this process is to enhance and extract critical signals from these complex datasets, enabling deeper insights. 
Image reconstruction methods recover high-quality images from incomplete or noisy data, while image enhancement techniques focus on improving clarity and contrast. For instance, preprocessing MRI data typically involves converting raw images to the Neuroimaging Informatics Technology Initiative (NIfTI) format, correcting for brightness inconsistencies and physiological noise, and isolating brain tissue from surrounding structures. 
Image registration aligns images across different time points or subjects to a standardized atlas, accounting for anatomical differences between individuals to ensure accurate comparisons. 
Image segmentation further partitions the images into meaningful regions, such as specific brain structures. \jianrevise{Partial volume effects occur when a single voxel contains a mixture of different tissue types, causing the measured signal to represent an average rather than a single homogeneous tissue class. This can blur boundaries between regions and bias tissue classification and quantitative measurements.}
Additionally, structural and functional neuroimaging data are often smoothed using a 3D Gaussian kernel, trading spatial resolution for increased statistical power. 
Analytical techniques vary, ranging from traditional 3D volume analysis to more advanced 2D cortical surface mesh approaches, which enhance visualization and improve analysis accuracy. For a more in-depth introduction to these processes, resources such as \citet{Lindquist2008,Jenkinson2017} provide an excellent starting point for beginners.

\jianrevise{The k-space sampling mechanism of MRI has important statistical consequences for image processing and analysis. Because artifacts arising from brief temporal events may be localized in k-space but propagate globally in image space after reconstruction, standard assumptions of spatially localized or independent noise can be violated. This motivates image reconstruction and enhancement methods that explicitly model the acquisition processes, leverage sparsity or low-rank structure in k-space, or jointly address motion and noise during reconstruction, rather than treating artifacts solely as post hoc image-space nuisances.} 

\section{Case Study 1: Brain Development from Birth to 20 Years}

\subsection{ Early Brain Development}
 
The development of the human brain across the first two decades of life can be divided into four key stages:
{\bf Stage I (0-2 years):} 
Rapid brain growth occurs, reaching 80\% of adult size~\citep{Knickmeyer2008,gilmore2018imaging}. This period is marked by sensory and motor development, increased synaptic formation, and extensive myelination in sensory and motor areas. 
{\bf  
Stage II (3-5 years):} 
Significant improvements in language, emotional control, and social skills occur, with development in the cerebral cortex and the beginning of synaptic pruning to enhance neural efficiency \citep{Fox2010}. 
{\bf  
Stage III (6-12 years):} 
Advances in logical thinking and problem-solving occur, with maturation in the frontal lobes, crucial for reasoning and decision-making. Synaptic pruning continues to optimize brain function \citep{Nelson2007,Giedd1999}. 
{\bf  
Stage IV (13-20 years):}  
Major changes in the frontal cortex enhance executive functions like planning and impulse control~\citep{Blakemore2006,Steinberg2005}. The limbic system matures, increasing emotional processing and sensitivity to social cues.
These stages are critical for understanding normal brain maturation and the mechanisms behind brain disorders rooted in developmental abnormalities.


\subsection{Cutting-edge Neuroimaging Studies on Early Brain Development} \label{sec:EBD}

\jianrevise{
The Human Connectome Project (HCP) \citep{Glasser2016a, VanEssen2013, Ugurbil2013} transformed MRI-based neuroimaging by establishing rigorous acquisition and preprocessing standards, which in turn catalyzed a new generation of large-scale developmental studies. Building on the HCP framework, these initiatives aim to map the structural and functional connectome across the lifespan, from infancy through early adulthood. For example, HCP-D \citep{somerville2018lifespan} characterizes connectivity and cognition in over 2,300 participants aged 5--21, while the ABCD Study \citep{casey2018adolescent, karcher2021abcd} longitudinally follows about 12,000 children to disentangle how environmental, behavioral, and genetic factors jointly shape brain maturation.
To probe the earliest stages of brain organization, the Baby Connectome Project (BCP) \citep{howell2019unc} tracks roughly 500 infants from birth to age 5. More recently, the HBCD Study \cite{HBCD2021} extends this developmental window further upstream by following approximately 7,500 children from the prenatal period through early childhood to quantify the effects of early-life exposures. 
By adopting harmonized imaging protocols rooted in HCP standards, these efforts enable unprecedented cross-study comparability. Collectively, they provide a coherent framework for tracing human brain development from prenatal origins through adolescence.}

\subsection{Data Challenges} \label{Sec:Challenges}

Three data challenges impact the analysis of neuroimaging and associated data types in studies focusing on early brain development. 
{\bf (DC1). Motion Artifacts, \jianrevise{Measurment Noise and Confounding}:} Significant head motion and noise during scans  can severely affect image quality and measurement accuracy, leading to false positives \citep{kennedy2022reliability,somerville2018lifespan,howell2019unc,hagler2019image,kaplan2022filtering}. Children often produce large motion artifacts, resulting in suboptimal images and shorter acquisition protocols that reduce tissue contrast compared to adult studies. This impacts various metrics, including cortical thickness, gray-matter volume, task fMRI, diffusion-tensor imaging, and resting-state networks. For example, task-fMRI measurements in children show poor reliability, with average within-session reliability of 0.088 and between-session stability of 0.072 \cite{kennedy2022reliability}, making it difficult to study individual differences in behavior and psychopathology.

Researchers address head motion using both prospective and retrospective correction techniques \citep{somerville2018lifespan, howell2019unc, hagler2019image}. Programs like BCP, HCP-D, and ABCD \citep{somerville2018lifespan, howell2019unc, hagler2019image, zhang2019resting} incorporate advanced MRI acquisition protocols and processing pipelines to mitigate issues related to motion, distortion, and intensity inhomogeneity. These corrections are continually refined to improve longitudinal data analysis and accommodate technological advancements. Additionally, head motion introduces biases that degrade data quality, requiring methods like censoring degraded frames and correcting distortions for accurate anatomical alignment.
  
\jianrevise{Beyond measurement noise, confounding represents a fundamental challenge in developmental neuroimaging. Motion artifacts are not purely random but are often correlated with subject characteristics such as age, neurodevelopmental status, attention, or clinical symptoms, raising the possibility of common upstream causes that influence both head motion and brain development. As a result, na{\"i}ve treatment of motion as noise can induce bias when motion-related variability is systematically associated with outcomes of interest. Moreover, commonly used remedies, such as scrubbing high-motion volumes or excluding participants with excessive motion, can unintentionally alter the target estimand by restricting inference to subpopulations with low motion, rather than the full population of children under study. Recent work has emphasized that such practices may compromise interpretability and comparability across studies, particularly in developmental cohorts \citep{peverill2025balancing}. These issues highlight the need for statistical methods that explicitly account for confounding mechanisms, rather than treating motion solely as an exogenous nuisance variable.}

 {\bf (DC2). Developmental Changes:}  
Dramatic changes in brain size, structure, and function \citep{Knickmeyer2008, Holland2014, gilmore2018imaging}  complicate data analysis. Smaller head sizes, partial volume effects, and ongoing myelination in younger brains result in low imaging contrast, signal-to-noise ratio, and spatial resolution \citep{Knickmeyer2008, Holland2014}. Infant brain MRI differs from adult MRI, showing reduced white-gray matter contrast between 0 to 9 months, large within-tissue intensity variations, and heterogeneous image appearances \citep{li2019computational}. Infant brain T1w and T2w images progress through three phases: (i) the infantile phase (0-3 months), where gray matter has higher intensity than white matter in T1w, offering better contrast in T2w; (ii) the isointense phase (5-9 months), where gray and white matter show minimal differentiation, complicating tissue segmentation and measurements; and (iii) the early adult-like phase (after 12 months), where T1w contrast begins to resemble that of adult brains.

Preprocessing infant MRI data requires specialized methods, as using adult pipelines can cause significant errors in downstream tasks. Over the past two decades, research has focused on infant-specific methods, particularly for MRI segmentation \citep{li2019computational, zhang2019resting, wang2023end, hagler2019image, wang2023ibeat}. Segmenting MRIs during the isointense phase remains especially challenging, and while current algorithms perform reasonably well, further improvements are needed, particularly in integrating brain topology, refining manual techniques, and assessing the uncertainty and generalizability of imaging methods and atlases in this complex context. 

{\bf (DC3).  Multimodal Data Integration:}  
 \jianrevise{Combining neuroimaging features with other data types, such as genetic and environmental information, from birth to age 20 presents unique analytical challenges (often referred to as data fusion or multimodal data integration in the literature). \citet{zhu2023statistical} identify several recurring themes in neuroimaging data analysis, including complex brain structure, spatiotemporal variation, high dimensionality, substantial subject heterogeneity, sampling bias, missing data, and complex causal relationships. Childhood and adolescence are periods of rapid brain growth and reorganization, which are critical for cognitive development \citep{van2022methodological}. However, neuroimaging protocols often vary across large developmental studies such as the BCP, HCP-D, and the ABCD study, complicating data harmonization and integration. Longitudinal studies also face high dropout rates, leading to incomplete observations and potential bias. In addition, there is growing interest in understanding how genetic and environmental factors, including familial risk, specific genetic variants, and prenatal exposures, jointly shape brain development \citep{gilbert2005genetic, zhou2024genetics, gao2019review, tooley2021environmental, miguel2019early}.
}

\subsection{Research Opportunities} \label{Sec:Opport}
 
These challenges present significant opportunities for statisticians, which we summarize below for brevity: 
\underline{(i) Experimental Design:} There is keen interest in optimizing study designs that accommodate the high variability and unique challenges of pediatric neuroimaging, such as motion artifacts from young children. This includes developing specialized experimental protocols that minimize movement and optimize data collection during periods when children are most cooperative. 
\underline{(ii) Uncertainty Quantification:} Implementing advanced statistical techniques to quantify the uncertainty in image processing algorithms is crucial, especially given the complications of motion artifacts and varying image quality. Effective uncertainty quantification enhances the credibility and reliability of findings in early brain development studies. 
\underline{(iii) Longitudinal Data Analysis:}  This requires the creation of statistical frameworks that can accommodate missing data, time-dependent changes, and intra-individual variability \citep{zhang2025sampling}, which are prevalent in long-term developmental studies. Accurately pinpointing critical developmental windows when the brain is particularly receptive to external factors is crucial, and statistical methods for event detection must be sensitive enough to discern subtle changes over time in the noisy environment of pediatric imaging data. 
\underline{(iv) Multimodal Analysis:} Integrating data from various imaging modalities (e.g., MRI, DTI, fMRI) with behavioral and genetic information can provide a more comprehensive view of brain development. This necessitates sophisticated multimodal data integration techniques that can handle data from different scales and modalities, allowing for deeper insights into developmental processes. 
\underline{(v) Causal Inference:} Developing statistical models to more accurately infer causal relationships in brain development data is critical, \jianrevise{given the need to address} confounding variables such as age, developmental stage, and environmental factors. Understanding how current interventions
impact development is essential to guide the development of
new interventions. 
By engaging with these statistical opportunities, researchers can enhance our understanding of brain development from infancy through adolescence, addressing both the technical data challenges and the complex biological questions involved. Such efforts will improve the design of interventions, deepen understanding of individual developmental trajectories, and potentially enable predictions of developmental outcomes based on early neuroimaging data.

\section{\jianrevise{Case Study 2: The Adult and Aging Brain}}

\subsection{ Stages of Adulthood}
 
Aging is a complex process involving biological, psychological, and social changes, and it varies significantly among individuals due to genetic, environmental, and lifestyle factors. 
\jianrevise{Differences in healthcare access and quality, both at the individual level and across generations, also contributes to variation in the rate and experience of aging.
}
While the aging process generally follows distinct stages, these factors lead to substantial variation in how each stage is experienced. Aging typically progresses through several stages~\citep{mousley2025topological}: 
 \textbf{ (I) Early adulthood (18 to late 30s):} 
 individuals
are often at their peak in health, strength, and agility.
 \textbf{(II)  Middle adulthood (40 to mid-60s):} Visible signs of aging (e.g., wrinkles, gray hair, and muscle loss) begin to appear, and chronic health conditions such as hypertension and arthritis become more common.
 \textbf{(III)  Late adulthood (mid-60s onward):} Physical changes, including reduced mobility, sensory decline, and increased susceptibility to illness, become more pronounced. Cognitive functions such as processing speed may decline, while accumulated knowledge remains relatively stable.
  \textbf{(IV) End of life}: Significant physical and cognitive decline may occur. 
  \jianrevise{Clinically, the
focus at this stage is primarily on symptom management and quality
of life rather than treatment of underlying disease.}

\subsection{Neuroimaging Studies of the Aging Brain}

Key datasets   include the Baltimore Longitudinal Study of Aging (BLSA), HCP Lifespan Connectome, Cambridge Centre for Aging and Neuroscience (Cam-CAN), UK Biobank (UKB), and Alzheimer's Disease Neuroimaging Initiative (ADNI).  
The {\bf BLSA}, the longest-running aging study in the US, began in 1958 and introduced brain imaging in 1994,  providing early longitudinal data on gray and white matter loss in older adults 
 \citep{resnick2003longitudinal}. The HCP Lifespan Connectome comprises three components: Development (HCP-D), Young Adult (HCP-YA), and Aging (HCP-A).
The {\bf HCP-D} was described in Section \ref{sec:EBD}. 
The {\bf HCP-YA} includes 1,200 subjects aged 21-35 years old, with rs-fMRI and task-based fMRI data from various cognitive tasks, plus test-retest and 7T movie-watching fMRI data on subsets of subjects. The {\bf   HCP-A} covers 725 subjects aged 36-100  years old, with similar fMRI data and arterial spin labeling (ASL) imaging. Cam-CAN involves 700 subjects aged 18-87  years old, with both rs-fMRI, task-based fMRI, and movie-watching data~\citep{Shafto2014CamCANProtocol,Taylor2017CamCANRepository}.
The {\bf UKB}  project includes brain imaging from a large cohort of middle-to-older adults (currently $n=100,000$), supporting research on brain aging. The {\bf ADNI} focuses on neuroimaging in Alzheimer's patients, those with mild cognitive impairment, and healthy controls, providing longitudinal data critical for studying neurodegenerative disease progression~\citep{Mueller2005ADNI,Weiner2010ADNIProgress}..
These datasets, both cross-sectional and longitudinal, are vital for tracking brain structure and function across the lifespan. Longitudinal studies are especially valuable for correlating brain changes with cognitive performance and identifying neural markers of age-related cognitive decline and dementia risk. Compared to cross-sectional data, longitudinal imaging is more effective for understanding individual trajectories of brain and cognitive aging. These studies also examine how lifestyle (e.g., physical activity, cognitive engagement, social interaction) and genetic factors influence brain resilience to age-related pathology.

\subsection{Data Challenges and Research Opportunities}

Analyzing neuroimaging data from aging populations presents several challenges.  Challenges (DC1)-(DC3) and the research opportunities outlined in  Section \ref{Sec:Challenges} carry-over to this cohort. Here we highlight several additional items related to DC2 and DC3 specific to studying the adult brain. 

{\bf (DC2) Developmental Changes in Aging}. 
 As humans age, the brain undergoes complex structural and functional changes, including 
 \jianrevise{cortical thinning, ventricular enlargement, and the appearance of white matter hyperintensities on T2-weighted and FLAIR MRI, which can reflect underlying structural pathology such as demyelination, axonal loss, and small-vessel ischemic damage ~\citep{Fazekas1987WMH,Wardlaw2013STRIVE}.}
 These changes are associated with declines in cognitive abilities, such as memory and executive function, which are typical of healthy aging. However, they can also indicate more severe cognitive decline, sensory impairments (e.g., hearing or vision loss), and an increased risk of neurodegenerative diseases like dementia. Aging affects individuals differently, with variations in brain structure and function influencing whether one experiences healthy aging or accelerated
decline.  Neurodegenerative diseases such as Alzheimer’s and Parkinson’s often overlap with typical aging, complicating the differentiation between disease-related and age-related changes in neuroimaging data.  
This individual variability makes it challenging to distinguish between normal aging and pathology.

Neuroimaging techniques are essential in aging research, offering several key applications. First, identifying neuroimaging markers of the biological processes underlying aging can help detect individuals at higher risk of age-related impairments. Second, neuroimaging biomarkers aid in the early detection of neurodegeneration and tracking disease progression, enabling timely diagnosis and intervention. 
\jianrevise{Third, these techniques provide critical insights into brain plasticity across the lifespan by characterizing how brain structure, function, and connectivity change in response to environmental factors and therapeutic or behavioral interventions.
}
Additionally, adjustments to neuropsychological assessments, imaging protocols, and experimental designs are necessary to ensure high-quality data collection in aging-related studies.

Detecting the subtle effects of aging, interventions, and disease on the brain requires large-scale studies \jianrevise{with sufficient statistical power to reliably characterize population-level trends and individual variability}. Recent work by \cite{cole2017predicting, cole2018brain} introduced an MRI-based biomarker to estimate a person's "brain age." The focus is on the brain age gap, the difference between brain age and chronological age, where a larger gap is indicative of poorer brain health. Additionally,  \cite{bethlehem2022brain} developed interactive brain charts using MRI data from over 100,000 participants to benchmark neurodevelopmental milestones and growth trajectories (\url{http://www.brainchart.io/}). The ENIGMA Lifespan working group analyzed data from 15,000 participants to map age-related declines in cortical thickness and explore genetic and environmental influences \citep{frangou2022cortical, dima2022subcortical, wierenga2022greater}. 
Using fMRI,  researchers have also revealed age-related changes in neural activity and functional connectivity, particularly in cognitive processes like memory, attention, and executive function. Resting-state fMRI often shows increased regional brain activation in older adults, possibly reflecting compensatory recruitment,often accompanied by reduced functional connectivity, which affects major brain networks, which may contribute to cognitive decline \citep{sala2015reorganization, geerligs2015brain}.

 {\bf (DC3) Integrating Multimodal Data in Aging}. Aging populations vary in genetics, lifestyle, and health, complicating statistical analyses. To reduce heterogeneity, it is crucial to stratify participants by relevant demographic and clinical variables. Advanced statistical methods are also needed to manage the complexities of neuroimaging data, including multiple comparisons, confounding factors, and intricate data structures. Ensuring the reliability, reproducibility, and accurate interpretation of findings, particularly in the context of age-related changes, is vital for advancing our understanding of the aging brain. 
Longitudinal data collection is essential for capturing changes over time, but faces challenges such as higher attrition rates due to health issues, relocation, or death. Effectively addressing missing data in these studies is critical. When managed well, these efforts can provide valuable insights into the neurobiology of aging and help develop strategies to promote healthy brain aging and prevent cognitive decline. 
Addressing these challenges will require interdisciplinary collaboration, methodological innovation, and a focused consideration of the unique characteristics of aging populations.

\section{Case Study 3: Neurodegeneration and Neuropsychiatric Disorders}

\subsection{The Human Brain and Neurodegeneration and Neuropsychiatric Disorders}

Neurodegenerative and Neuropsychiatric disorders both stem from core disruptions in brain structure and function. Although these conditions present differently and are often studied separately due to their unique characteristics, it is essential to understand their connections. 

Neuropsychiatric disorders mainly affect human cognition, emotions, and behaviors. Common examples include major depressive disorder, schizophrenia, and bipolar disorder. In major depressive disorder, structural and functional changes in the prefrontal cortex and hippocampus (key areas for mood regulation and memory) are often observed~\citep{otte2016major}. Reduced hippocampal volume, linked to prolonged stress, may worsen depressive symptoms such as emotional dysregulation and cognitive impairment.  Schizophrenia involves brain  volume reduction, particularly in the frontal and temporal lobes, which impacts cognition and perception. This strucural loss is coupled with  disruptions in dopamine and glutamate systems, contributing to halmark symptoms like hallucinations, delusions and impaired cognitive functions~\citep{mccutcheon2020schizophrenia}. Bipolar disorder is characterized by structural and functional changes in brain regions such as the amygdala and prefrontal cortex~\citep{grande2016bipolar}, which regulate emotion. Neuropsychiatric disorders result from both structural brain abnormalities and neurotransmitter imbalances. Disruptions in brain regions and neural pathways, combined with altered neurotransmitter systems, contribute to the cognitive and emotional symptoms characteristic of these conditions.  

Neurodegenerative disorders are characterized by the gradual and irreversible loss of neurons and their functions, often resulting in cell death. This progressive degeneration disrupts key neural circuits and impairs essential brain functions. Among the most well-known neurodegenerative disorders are Alzheimer's disease, Parkinson's disease, and huntingtin's disease, each with distinct pathological mechanisms.  Alzheimer's disease is primarily associated with the accumulation of amyloid-beta plaques and tau protein tangles within the brain. These abnormal protein deposits disrupt neural communication and trigger inflammation, leading to the gradual loss of neurons and synapses, particularly in the cerebral cortex and subcortical areas. This neuronal death contributes to the hallmark symptoms of Alzheimer's, including memory loss, disorientation, and cognitive decline as the disease progresses~\citep{scheltens2021alzheimer}. Parkinson's disease is distinguished by the degeneration of dopamine-producing neurons in the substantia nigra, a key region of the midbrain involved in motor control. The depletion of dopamine disrupts signaling between the brain and muscles, leading to motor symptoms such as tremors, stiffness, and bradykinesia (slowness of movement). As Parkinson's disease advances, non-motor symptoms, including cognitive impairment and mood disturbances, may emerge, reflecting the broader impact of neuronal loss on brain function~\citep{bloem2021parkinson}. Huntington's disease, in contrast,  is a genetic disorder caused by mutations in the HTT gene, leading to the production of an abnormal form of the huntinghtin protein. This mutated protein accumulates within neurons, particularly in brain regions like the striatum, which plays a crucial role in movement, mood regulation, and cognition. Over time, the accumulation of abnormal proteins leads to widespread neuronal death, manifesting in symptoms such as uncontrolled movements (chorea), mood swings, and cognitive decline~\citep{walker2007huntington}. These disorders involve progressive neuronal loss that leads to the deterioration of cognitive and physical abilities over time.

Although traditionally studied separately, neuropsychiatric and neurodegenerative disorders share several overlapping features and connections. Both can involve neurotransmitter imbalances, inflammation, oxidative stress, and issues with protein processing. Certain genetic factors may increase susceptibility to both types of disorders, indicating shared molecular pathways. Cognitive deficits, mood disturbances, and behavioral changes can occur in both, even though their primary symptoms and progression differ. Gaining insight into the biochemical and structural alterations in the brain can help guide treatment approaches, including medications targeting specific neurotransmitter systems and therapies focused on neuroprotection and regeneration.

Research exploring the links between these disorders is rapidly expanding, driven by advances in neuroimaging and molecular biology. Techniques like MRI, PET, fMRI, and EEG have provided deeper insights into their complex interrelationships. For instance, studies using fMRI and EEG have identified dysfunctional connectivity in the default mode network and cognitive control circuits in neuropsychiatric disorders, offering new insights into underlying mechanisms and potential targets for intervention~ \citep{Greicius2007,Buckner2008DMN,Cole2014}.  The integration of these findings with emerging brain-stimulation techniques, such as transcranial magnetic stimulation (TMS) and deep brain stimulation (DBS), holds promise for developing personalized therapies that are precise and effective. These approaches not only enhance our understanding of the disorders but also improve diagnostic accuracy and treatment strategies, ultimately aiming to improve the quality of life for those affected.

Understanding the connections between the human brain and these disorders is essential. The profound impact that various brain-related disorders have on individuals, families, and society underscores the ethical imperative to study and address them. By delving into the neural underpinnings of these conditions, researchers can develop strategies to alleviate suffering, reduce associated burdens, and promote mental health across populations.

\begin{figure}[htbp]\label{newfigure3}
\begin{center}
\includegraphics[width=1\textwidth]{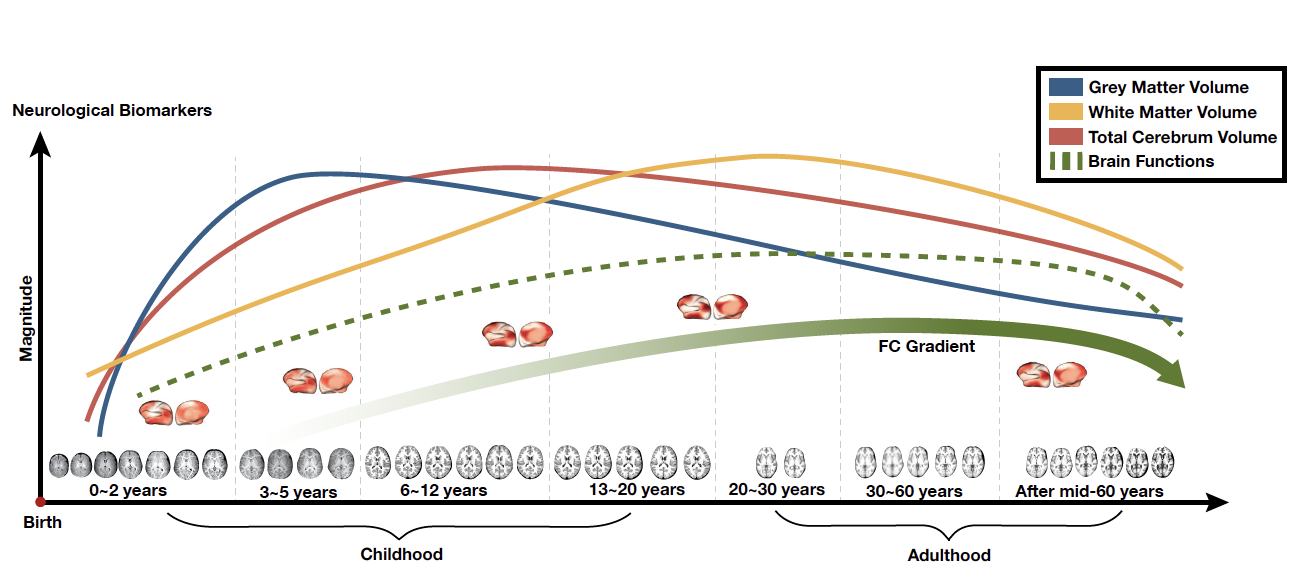}\\
\includegraphics[width=1\textwidth]{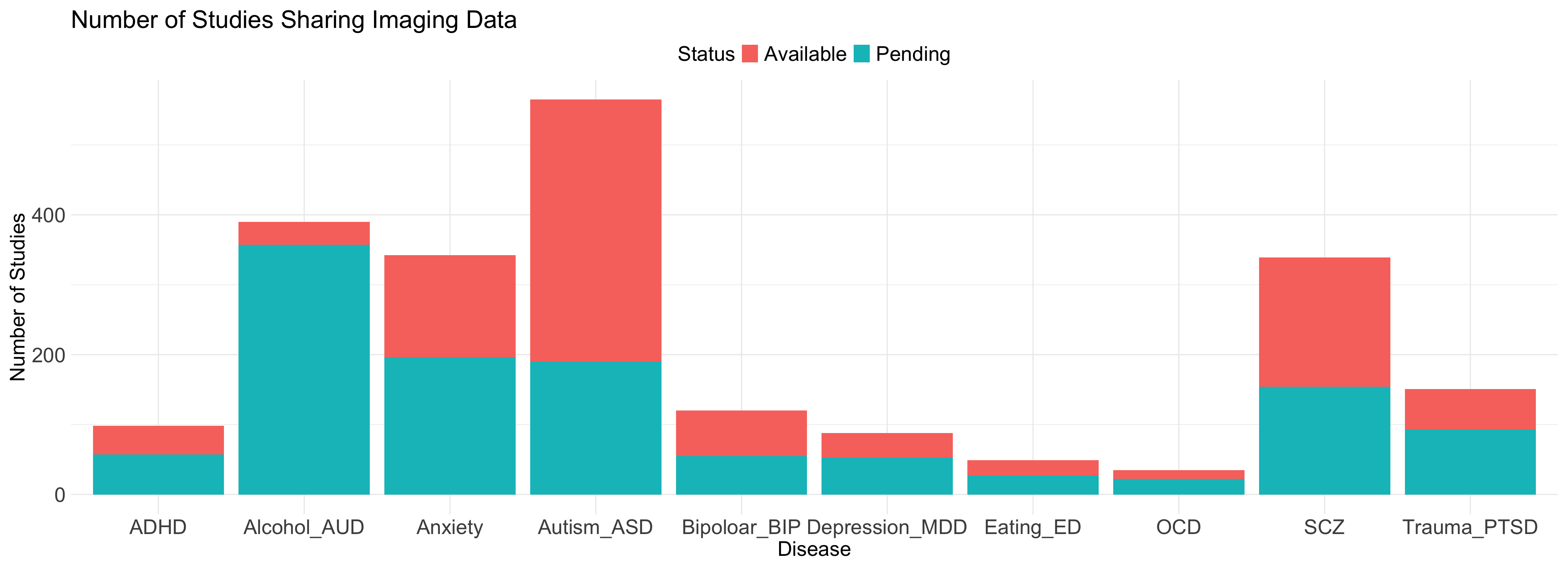}\\
\end{center}
\caption{Top: Brain development across lifespan partially extracted from \citet{bethlehem2022brain}. Bottom: Number of neuroimaging studies sharing neuroimaging data 
 by disease and submission status in the National Institute of Mental Health Data Archive (NDA).}
\end{figure}

\subsection{Neuroimaging Studies}
The National Institute of Mental Health Data Archive (NDA) (\url{https://nda.nih.gov/}) is a vital resource for statisticians focused on neuroimaging data analysis in mental health research. Originally established to consolidate data from projects such as the National Database for Autism Research (NDAR) and the NIH Pediatric MRI Repository, the NDA has since evolved into a comprehensive platform that integrates diverse datasets into a harmonized database. Figure~\ref{newfigure3} (bottom) illustrates the number of biomedical studies contributing neuroimaging data, categorized by psychiatric disorder and submission status. 
For statisticians, the NDA provides access to clinical, neuroimaging, phenotypic, and genomic data from hundreds of thousands of participants. Statisticians can leverage this rich resource to access datasets, perform sophisticated data analyses, develop advanced models, identify patterns linked to psychiatric disorders, and support the creation of targeted interventions. With structured access through permission groups, the NDA's strong commitment to data sharing makes it an indispensable tool for advancing mental health research.

We briefly review two major initiatives for neurodegenerative disorders: the Alzheimer's Disease Neuroimaging Initiative (ADNI) and the Parkinson's Progression Markers Initiative (PPMI). 
The {\bf ADNI}~\citep{mueller2005alzheimer} is a pioneer longitudinal study in Alzheimer's disease (AD) research, designed to integrate multiple modalities, including MRI \citep{jack2008alzheimer}, PET \citep{jagust2010alzheimer}, biological markers ~\citep{kang2015alzheimer}, and detailed clinical and neuropsychological assessments~\citep{park2012confirmatory}. This approach aims to map the progression from normal cognition to mild cognitive impairment (MCI) and ultimately to Alzheimer's disease. By collecting data across different disease stages, ADNI seeks to identify reliable amyloid-beta-tau-Neurodegeneration biomarkers and predictors of AD progression. The {\bf   PPMI}  is a landmark multi-center observational study focused on identifying and validating biomarkers for Parkinson's disease (PD). It collects clinical, imaging, and biosample data from over 1,400 PD patients, 400 healthy controls, and various at-risk individuals, such as those with REM sleep behavior disorder (RBD), genetic mutations associated with PD, and first-degree relatives of PD patients. PPMI remains an essential resource in the fight against Parkinson's disease.

We briefly review four key initiatives for neuropsychiatric disorders.
The Autism Brain Imaging Data Exchange ({\bf ABIDE}), launched in 2012, advances understanding of autism spectrum disorder (ASD) by openly sharing neuroimaging data from over 1,000 individuals with ASD and 1,000 controls across multiple international sites. Using advanced imaging techniques such as fMRI and sMRI, ABIDE enables researchers to investigate brain connectivity patterns and potential biomarkers, significantly contributing to reproducibility and collaboration in autism research. 
The {\bf IMAGEN} study, initiated in 2007, is a large-scale European project that explores the biological and environmental factors influencing mental health and risk-taking behavior in adolescence. With over 2,000 participants, it integrates neuroimaging, genetic, and behavioral data to track brain development and its association with cognitive and emotional functions, providing insights into the onset of psychiatric disorders. Its longitudinal design allows for a comprehensive understanding of adolescent brain development. 
The Bipolar \& Schizophrenia Consortium for Parsing Intermediate Phenotypes ({\bf B-SNIP 1}), launched in 2008, focuses on uncovering the neurobiological foundations of psychotic disorders, including bipolar disorder and schizophrenia. The study, involving over 1,000 participants, uses a variety of imaging techniques like sMRI, fMRI, and diffusion tensor imaging (DTI) to identify intermediate phenotypes, which may serve as biomarkers for diagnosis and treatment. 
The Sleep to Reduce Incident Depression Effectively ({\bf STRIDE}) study investigates how sleep interventions can prevent depression. It explores the link between sleep disturbances and depression by evaluating sleep quality improvements through cognitive-behavioral therapy for insomnia (CBT-I) and other interventions. The study's neuroimaging data, collected from over 600 participants, helps clarify the neural mechanisms connecting sleep and mood, guiding preventive mental health strategies.

\subsection{Data Challenges and Research Opportunities}

\jianrevise{Analyzing neuroimaging data within neurodegenerative and neuropsychiatric populations introduces a layer of complexity beyond that of healthy aging studies. While core challenges (DC1–DC3) remain relevant, they are fundamentally altered by pathological heterogeneity and non-linear progression. 
We first examine how disease-specific physiological noise and structural variability intensify (DC2) and (DC3). We then introduce a fourth challenge (DC4) unique to this context: the statistical decoupling of pathological evolution from the underlying processes of normal aging.}

{\bf (DC2). Disturbed Developmental Changes:}   
The neuroimaging data and methods used in Case Studies 1 and 2 allow us to map life-span trajectories of brain structure and function in healthy individuals, enabling the creation of "growth charts" for normal brain development. These charts can serve as valuable references for studying and comparing neuroimaging data from diseased populations.  

\jianrevise{Building on these normative references, the analysis of neurodegenerative and neuropsychiatric diseases requires statistical methods capable of extracting disease-related signals from high-dimensional neuroimaging data. Each imaging modality offers a distinct view of brain structure and/or function, but the resulting data are characterized by an extremely large number of voxels and complex spatial dependence. Advanced statistical and machine learning  techniques, including variable selection~\citep{fan2010selective, wang2017generalized,kang2018scalar,jin2023bayesian,liu2024robust,wu2025bayesian}, functional data analysis~\citep{ramsay2005functional,morris2015functional}, dimension reduction~\citep{van2009dimensionality,kingma2019introduction}, graphical model~\citep{huang2010learning, kang2016depression, kundu2016semiparametric, 9763540}, 
deep neural network model \citep{parisot2018disease,velickovic2017graph,lian2018hierarchical},  
and manifold learning~\citep{van2023alzheimer, he2025scalable, Li2026BayesianSoN}, are therefore essential for identifying potential biomarkers and building predictive models of disease progression, such as in Alzheimer’s disease~\citep{mueller2005ways}.
Longitudinal neuroimaging studies further allow disease progression to be characterized over time, but introduce additional statistical challenges, including modeling temporal dynamics, handling informative dropout and missing data, addressing time-dependent confounding, and performing inference under irregular follow-up. Methodological frameworks such as mixed-effects models~\citep{bernal2013statistical, cobigo2022detection,zhu2019fmem}, spatial process models~\citep{ziegler2014individualized}, functional regression~\citep{goldsmith2012longitudinal, li2022regression},  time-series analysis~\citep{yang2020disentangling}, and survival analysis~\citep{sorensen2019prognosis} play a central role in addressing these issues and in linking longitudinal imaging changes to clinical outcomes and treatment effects~\citep{donohue2014estimating}.}

{\bf (DC3). Causal Multimodal Data Integration:} 
Multimodal data integration is crucial for addressing key scientific questions related to neuropsychiatric and neurodegenerative disorders. These questions include  uncovering causal relationships between genetics, brain function, health factors, and diseases,   developing targeted treatments that address brain structure and/or function associated with these diseases, and understanding how the brain mediates the relationship between therapies or drugs and clinical outcomes. Tackling these questions presents significant statistical challenges in experimental design, data collection, multimodal data integration methods, causal discovery, causal modeling, and causality validation~\citep{guo2022spatial, yu2022mapping, chang2023statistical, li2019spatially}.

{\bf (DC4). Heterogeneous Disease Patterns:}   
Heterogeneous disease patterns in brain disorders present significant challenges in both clinical care and research. Neuroimaging plays a crucial role in detecting disease-specific markers, helping clinicians and researchers better understand underlying pathologies \citep{doi2007computer}. However, substantial spatial and temporal variability exists in the affected brain regions across patients, complicating the development of standardized approaches for diagnosis and treatment \citep{liu2021statistical,EBRAHIMIGHAHNAVIEH2020105242,liu2025hcdpd,huang2022dadp}. 
Spatial heterogeneity is evident in the variability of abnormal brain regions, with disease patterns being consistent only within a small subset of patients.   Similarly, temporal heterogeneity occurs as disease progression—such as brain degeneration or loss of neural connectivity—varies significantly over time. 
Understanding this imaging heterogeneity is crucial for advancing precision medicine, where treatments can be tailored to individual disease presentations. Identifying specific biomarkers and imaging patterns associated with different subtypes allows clinicians to improve the accuracy of diagnosis, prognosis, and treatment planning. Addressing this variability is key to enhancing prevention, diagnosis, treatment, and management of neuropsychiatric and neurodegenerative diseases.

{\it New Research Opportunities:}
Statisticians play a critical role in neuroimaging, particularly in validating imaging biomarkers. These biomarkers are quantitative indicators derived from neuroimaging data that reflect normal or pathological biological processes or responses to therapeutic interventions. Rigorous statistical testing ensures that identified biomarkers are not only statistically significant but also robust, reproducible, and clinically relevant. This involves applying advanced statistical techniques to assess their reliability and validity across different populations and settings.

Validation methods such as cross-validation and bootstrap techniques are commonly used to estimate model performance and prevent overfitting. Cross-validation partitions data into subsets, training the model on some subsets while validating it on others, then averaging the results to obtain an overall performance metric. Bootstrap methods involve repeatedly resampling data with replacement to create multiple simulated samples, allowing estimation of the sampling distribution of a statistic. External validation using independent datasets is also crucial, as it tests the generalizability of biomarkers to new, unseen data—essential for clinical application~\citep{varoquaux2017assessing, tibshirani1993introduction}.

By ensuring statistical rigor in the validation process, statisticians establish the credibility of these biomarkers, which can significantly impact the diagnosis and treatment of neuropsychiatric and neurodegenerative diseases. Reliable biomarkers aid in early detection, monitor disease progression, and evaluate the effectiveness of therapeutic interventions. They also facilitate personalized medicine by identifying patient subgroups that may respond differently to specific treatments, ultimately improving patient outcomes.

The ethical handling of sensitive neuroimaging data presents significant opportunities for statisticians to ensure privacy and data protection. Statisticians must implement secure storage, encrypted transmission, and access control measures to safeguard personal information. Ethical challenges begin with obtaining informed consent, clearly communicating how data will be used and protected. Statisticians play a key role in ensuring data anonymization through techniques like de-identification and pseudonymization.

In addition, statisticians help balance open science with confidentiality by adhering to data sharing guidelines, institutional review board (IRB) approvals, and data use agreements. By upholding these ethical standards, statisticians contribute to responsible research that fosters trust, protects participants, and advances scientific knowledge.

\section{Case Study 4: Brain Encoding-Decoding Research}

\subsection{Encoding-Decoding Tasks}

Encoding-decoding is a fundamental problem in system, cognitive, and computational neuroscience. It centers around understanding how the brain encodes information from the external world, presented as stimuli, into neural signals, and how these signals can be decoded to infer and reconstruct the original stimulus or mental state. \jianrevise{Unlike the preceding case studies, which are organized around populations or disease cohorts, encoding–decoding research is organized around computational tasks, providing a complementary methodological perspective on statistical challenges in neuroimaging. } The problem traces back to pioneering studies like Edgar Adrian's work in the 1920s on the electrical activity of the nervous system, and has received revived interest in recent years, partially due to the rapid advances of deep learning techniques. See \citet{Kriegeskorte2019} and \citet{Cao2021} for some reviews.

Applications of the encoding-decoding framework are vast and impactful. A major application area is brain-computer interface (BCI), which allows for the control of external devices with neural signals, and offers new avenues for individuals with motor disabilities \citep{Saha2021}. Another application area is the diagnosis and treatment of neurological disorders, where understanding the neural correlates of diseases like epilepsy and Alzheimer's disease can lead to better therapeutic strategies \citep{Greicius2003}. Last but not least, the encoding-decoding research significantly enhances our understanding of how thoughts, memories, and emotions are represented in the brain.

\jianrevise{The major imaging modalities used in encoding and decoding research: fMRI, EEG, and MEG, offer complementary strengths that shape the types of representations and models that can be learned. fMRI, with its high spatial resolution, is particularly well suited for encoding and decoding distributed cortical representations, enabling voxel-wise or region-level models that link complex stimuli or cognitive states to spatial patterns of brain activity. EEG, including invasive recordings such as ECoG and non-invasive scalp measurements, provides high temporal resolution that supports decoding of fast neural dynamics and time-locked responses, making it especially valuable for studying sequential processing, attention, and real-time brain–computer interfaces. MEG similarly enables millisecond-scale temporal decoding while offering improved spatial localization compared to EEG, facilitating models that capture both temporal dynamics and spatial organization of neural responses. Together, these modalities motivate different modeling choices in encoding and decoding frameworks, ranging from spatially structured regressions to dynamical and sequence-based models, depending on the resolution and signal characteristics of the data.}

Encoding-decoding research presents numerous intriguing and challenging questions that demand the development of new statistical and machine learning methods. The encoding task focuses on mapping stimuli to patterns of neural activity across brain regions. This involves identifying which regions are activated by specific stimuli, understanding the temporal dynamics of neural responses, and determining how different types of stimuli, such as images and language, are represented in the brain. It requires a detailed analysis of spatial and temporal brain activity to uncover how sensory input is processed across cortical and subcortical regions. 
On the other hand, the decoding task seeks to interpret neural activity patterns to predict stimuli or subject responses. This includes reconstructing external stimuli from brain activity, predicting behavior in response to stimuli—such as movements or decisions—and identifying cognitive states like attention, memory recall, or emotions. Decoding requires advanced statistical and machine learning methods capable of handling complex, noisy brain data and reliably distinguishing neural patterns associated with specific stimuli or responses.

\subsection{Neuroimaging Studies}

Several publicly available benchmark datasets are used for encoding-decoding tasks, often categorized by neuroimaging modality and learning task. Two common examples are EEG data with text and fMRI data with natural images.   The {\bf ZuCo}  dataset \citep{ZuCo1, ZuCo2}, short for Zurich Cognitive Language Processing Corpus, is a valuable resource for studying the cognitive and neural mechanisms of natural language processing. 
\jianrevise{It combines eye-tracking and EEG data from 12 native English-speaking participants engaged in a range of well-defined language tasks, including sentence reading, semantic processing, and natural reading paradigms.} 
This multimodal approach enables an integrated analysis of how visual processing and brain activity are coordinated during reading. 
ZuCo 1.0 includes 400 sentences per subject, while ZuCo 2.0 adds sessions involving emotional and task-free reading. EEG data is collected using a 128-channel system at 500Hz, with 105 channels used after noise filtering. \jianrevise{This relatively high channel density provides finer spatial sampling than conventional low-density EEG systems, which is advantageous for encoding–decoding analyses that aim to capture distributed neural representations during language processing.} The dataset's rich multimodal nature provides a unique window into real-time language comprehension and cognitive processing dynamics.

The {\bf GOD} dataset \citep{Horikawa2017}, short for Generic Object Decoding, is a comprehensive neuroimaging and behavioral dataset designed to study human visual cognition and neural representations of visual stimuli. It includes high-resolution fMRI data and high-quality images of objects and landscapes viewed by participants during scanning, allowing researchers to examine brain regions involved in cognitive tasks like object recognition and spatial navigation. 
The dataset covers a broad range of stimuli, including natural scenes and everyday objects, and records behavioral responses, such as reaction times and accuracy, \jianrevise{which enables the study
of associations between neural data and cognitive performance.}
With hundreds of subjects viewing thousands of images, the GOD dataset is a valuable resource for understanding how the brain processes diverse visual environments.

The Natural Scenes Dataset ({\bf NSD}) \citep[NSD]{Allen2022} is designed to explore how the brain processes natural images, featuring high-resolution 7T fMRI data and a vast array of stimuli from the Microsoft Common Objects in Context dataset \citep[CoCo]{CoCo}. Eight subjects viewed thousands of natural images for 3 seconds each while engaged in a continuous recognition task, indicating whether they had seen the image before. 
The dataset’s longitudinal design, with multiple scanning sessions, allows for studying the consistency and variability of neural responses over time. It also includes behavioral data, linking brain activity to cognitive functions like memory and perception. NSD offers valuable insights into the neural mechanisms of visual perception and is a key resource for advancing our understanding of how the brain interprets complex natural environments.

\subsection{Data Challenges and Research Opportunities}

\jianrevise{
Encoding--decoding tasks share several core statistical challenges with the population-based case studies discussed earlier, particularly high dimensionality, limited sample sizes, and substantial measurement noise (DC1--DC3). At the same time, encoding--decoding problems introduce additional challenges that are specific to task-driven, stimulus–response modeling and relatively small sample size.
}

\jianrevise{
\noindent\textbf{(DC1) Measurement Noise and Trial-to-Trial Variability.}
Encoding--decoding analyses are especially sensitive to measurement noise and trial-to-trial variability. recorded neural signals can vary substantially even under repeated presentations of identical stimuli, due to intrinsic biological variability as well as technical limitations of acquisition devices. This issue is particularly pronounced in non-invasive EEG studies, where signals must propagate through the scalp and skull, resulting in low spatial resolution and spatially mixed measurements. Consequently, observed signals often represent superpositions of multiple neural sources, complicating stimulus reconstruction and decoding accuracy. These challenges underscore the need for advanced statistical methods that explicitly model noise, uncertainty, and latent neural structure~\citep{ma2022bayesian, zhao2025bayesian}.}

\jianrevise{\noindent\textbf{(DC2) Nonlinear and Time-Dependent Neural Dynamics.}
The relationship between neural activity and represented stimuli or behaviors is often highly nonlinear and time-dependent, reflecting complex neural dynamics that vary across experimental conditions and cognitive states. Capturing such dynamics substantially increases model complexity and exacerbates challenges associated with high dimensionality and limited training data. Deep learning methods have therefore become prominent tools in encoding--decoding research. However, their application introduces additional task-specific challenges, including substantial computational burden, data inefficiency, and difficulties in ensuring interpretability and biological plausibility, motivating the development of novel statistical and hybrid modeling approaches~\citep{ma2023bayesian, zhou2024sequential, Zhao2025BayesianRLBCI,chen2018neural}.}

\jianrevise{
\noindent \textbf{(DC5) Small sample sizes.} Limited data exacerbate overfitting, instability, and poor generalization, while simultaneously increasing the importance of principled uncertainty quantification and model interpretability. This setting motivates the development of methods that can efficiently borrow strength across subjects, tasks, modalities, and studies through hierarchical modeling, transfer learning, and structured regularization. In particular, Bayesian and hybrid statistical–machine learning frameworks provide a natural mechanism for incorporating prior information, leveraging external or population-level data, and adapting pre-trained representations to task-specific settings. Viewed from this perspective, the prevalence of small-sample studies is not merely a limitation, but a defining characteristic that motivates robust, data-efficient, and interpretable approaches to brain encoding–decoding analysis~\citep{ma2025bayesian}.
}

\section{Discussion}
\jianrevise{
This section synthesizes statistical opportunities across the four neuroimaging case studies, integrating domain-specific challenges into a unified statistical perspective. Rather than isolating statistical issues in a standalone discussion, we highlight how core opportunities---including uncertainty quantification, causal inference, longitudinal modeling, and scalable multimodal integration---recur across distinct scientific contexts. Table~\ref{tab:crosscutting} summarizes challenges that are shared across the four neuroimaging domains considered in this paper, as well as those that are specific to particular contexts. Although many challenges are common, their manifestations and methodological implications differ substantially across domains, giving rise to distinct yet interconnected opportunities for methodological development.
}

\begin{table}[htbp]
\centering
\caption{Cross-cutting and case-specific statistical challenges across neuroimaging domains.}
\label{tab:crosscutting}
\jianrevise{
\begin{tabular}{lcccc}
\hline
\textbf{Statistical Challenge} 
& \textbf{Development} 
& \textbf{Adult Aging} 
& \textbf{Disease} 
& \textbf{Encoding--Decoding} \\
\hline
High dimensionality & $\checkmark$ & $\checkmark$ & $\checkmark$ & $\checkmark$ \\
Measurement noise (DC1) & $\checkmark$ & $\checkmark$ & $\checkmark$ & $\checkmark$ \\
Subject heterogeneity (DC2) & $\checkmark$ & $\checkmark$ & $\checkmark$ & $\checkmark$ \\
Limited effective sample size (DC3) & $\checkmark$ & $\checkmark$ & $\checkmark$ & $\checkmark$ \\
Longitudinal missingness & $\checkmark$ & $\checkmark$ & $\checkmark$ & -- \\
Rapid anatomical change & $\checkmark$ & -- & -- & -- \\
Age-related structural decline & -- & $\checkmark$ & -- & -- \\
Pathological heterogeneity (DC4) & -- & -- & $\checkmark$ & -- \\
Temporal alignment and dynamics & -- & -- & -- & $\checkmark$ \\
Computational and real-time constraints & -- & -- & -- & $\checkmark$ \\
\hline
\end{tabular}
}
\end{table}

Neuroimaging offers a transformative frontier for statisticians, requiring sophisticated frameworks to address the ``curse of dimensionality'' and the complex structure of neural data \citep{marron2021object,zhu2023statistical}. Modern tasks---from dimensionality reduction and feature selection to multimodal integration---draw on a broad toolkit spanning manifold learning, object-oriented data analysis, and multivariate statistics. To improve portability across cohorts, scanners, and populations, researchers increasingly rely on transfer learning and domain adaptation, as well as ensemble and federated learning strategies \citep{guan2021domain,guan2024federated,shan2024merging,fang2024source}. In particular, as privacy constraints and institutional silos become more prominent, federated learning provides a practical pathway for large-scale multi-site collaboration without centralized data sharing.

Longitudinal neuroimaging studies introduce additional challenges due to irregular sampling, informative dropout, and pervasive missingness \citep{zhu2019fmem,zhang2025sampling}. Addressing these issues calls for robust statistical models that can capture nonlinear trajectories of development and decline, while properly accounting for measurement error and attrition. Beyond time, integrating multimodal information across imaging modalities, behavioral assessments, genetics, and clinical variables further complicates analysis \citep{baltruvsaitis2018multimodal,jiang2025computation}. Classical approaches such as canonical correlation analysis and joint factorization remain useful \citep{shu2020d,lock2013joint}, but increasingly must be combined with scalable regularization, uncertainty quantification, and principled validation to support modern high-dimensional settings.

Causal questions in neuroimaging go beyond association and require careful definition of estimands, adjustment for confounding, and sensitivity analyses under realistic data-generation mechanisms \citep{guo2022spatial, yu2022mapping, chang2023statistical, li2019spatially,yu2022mapping}. At the same time, ethical considerations, including privacy, consent, and governance for secondary use, must be addressed alongside methodological development.

A rapidly emerging opportunity is the development of \emph{foundation models} for neuroimaging processing and analysis \citep{cox2024brainsegfounder,zhang2025generalist,sun2025foundation,tian2024unigradicon,tak2026generalizable}, together with \emph{digital twin} brain frameworks for personalized simulation and forecasting \citep{lu2024imitating,xiong2023digital,guo2025ten}. Both directions promise transferable representations, improved cross-site harmonization, and data-efficient adaptation across tasks (e.g., segmentation, phenotyping, prognosis, and multimodal prediction). Realizing this promise, however, requires statistical foundations for rigorous benchmarking, uncertainty quantification and calibration, robustness to distribution shift, bias/fairness assessment, and interpretable links between learned representations and neurobiological mechanisms.

To support methodological innovation, a growing ecosystem of neuroimaging datasets and review resources is available. A curated index of useful datasets is provided at \href{https://statsupai.org/pages/neuroimaging2.html}{statsupai.org}, facilitating access to diverse imaging data for developing and validating new methods.
\jianrevise{In addition, \href{https://statsupai.org/pages/Neuroimaging_review_papers.html}{statsupai.org} curates links to comprehensive review papers in neuroimaging, directing readers to critical analyses of current methodologies and recent developments in neuroimaging analysis, some of which may require subscription or institutional access.}

\jianrevise{
Looking forward, the expanding availability of high-dimensional, longitudinal, and multimodal neuroimaging data underscores the need for specialized analytical tools, deeper interdisciplinary collaboration, and a sustained emphasis on reproducibility and open science. While advanced machine learning methods offer powerful modeling capabilities, their reliability depends on statistical foundations that ensure interpretability, validation, uncertainty quantification, and generalizability. Statisticians therefore play a central role in defining estimands, addressing confounding and missing data, and developing scalable inference frameworks that connect neuroimaging with personalized clinical information. These efforts are essential for translating methodological advances into trustworthy scientific insights and, ultimately, for advancing precision medicine and improving clinical outcomes.
}



\section*{Funding} Dr. Kang is partially supported by the National Institute of Health (NIH) grants R01DA048993 and R01MH105561 and the National Science Foundation (NSF) grant IIS2123777.  Dr. Nichols is partially supported by the NIH grant R01DA048993. Dr. Li is partially supported by the NIH grants R01AG061303, R01AG080043 and the NSF grant CIF-2102227. 
Dr. Lindquist  is partially supported by R01 EB026549 from the National Institute of Biomedical Imaging and Bioengineering and R01 MH129397 from the National Institute of Mental Health. Dr. Zhu is partially supported by  the NIH grants RF1AG082938, R01AG085581,  
 R01MH136055, and R01AR082684 
and the Gillings Innovation Laboratory on generative AI. The content is solely the responsibility of the authors and does not necessarily represent the official views of NIH and NSF.

\appendix
\section*{Accessibility and Primary References for Major Neuroimaging Datasets}

\setcounter{table}{0}
\renewcommand{\thetable}{A}

\noindent
\jianrevise{Table~\ref{tab:dataset_access} summarizes major neuroimaging datasets referenced in this review, including imaging modalities, target populations, access mechanisms, and representative citable references. Dataset availability and access policies may evolve over time.}

\begin{sidewaystable}[p]
\centering
\caption{Accessibility and representative citable references for neuroimaging datasets}
\label{tab:dataset_access}
{\color{black}
\begin{tabular}{p{3.0cm} p{3.8cm} p{4.2cm} p{3.2cm} p{5.5cm}}
\hline
\textbf{Dataset} & \textbf{Imaging Modalities} & \textbf{Population} & \textbf{Access Type} & \textbf{Representative Reference (DOI)} \\
\hline

HCP (HCP-D, HCP-A)
& sMRI, fMRI, dMRI
& Lifespan
& Open / Restricted
& \href{https://doi.org/10.1016/j.neuroimage.2013.05.041}{10.1016/j.neuroimage.2013.05.041} \\

ABCD
& sMRI, fMRI, dMRI
& Children (9--10 at baseline)
& Controlled (NDA)
& \href{https://doi.org/10.1016/j.dcn.2018.03.001}{10.1016/j.dcn.2018.03.001} \\

BCP
& sMRI, fMRI, dMRI
& Infants (0--5 years)
& Controlled
& \href{https://doi.org/10.1016/j.neuroimage.2019.01.041}{10.1016/j.neuroimage.2019.01.041} \\

HBCD
& sMRI, fMRI, dMRI
& Prenatal to early childhood
& Controlled
& \href{https://doi.org/10.1038/s41591-021-01643-0}{10.1038/s41591-021-01643-0} \\

UKB
& sMRI, fMRI, dMRI, genetics
& Middle-aged and older adults
& Application-based
& \href{https://doi.org/10.1038/s41586-018-0579-z}{10.1038/s41586-018-0579-z} \\

ADNI
& sMRI, PET, biomarkers
& Aging, MCI, Alzheimer's disease
& Controlled
& \href{https://doi.org/10.1016/j.neuroimage.2010.03.081}{10.1016/j.neuroimage.2010.03.081} \\

PPMI
& sMRI, PET, clinical measures
& Parkinson's disease
& Controlled
& \href{https://doi.org/10.1002/mds.25654}{10.1002/mds.25654} \\

ABIDE
& sMRI, fMRI
& Autism spectrum disorder
& Open
& \href{https://doi.org/10.1038/mp.2014.78}{10.1038/mp.2014.78} \\

NSD
& 7T fMRI
& Healthy adults
& Open
& \href{https://doi.org/10.1038/s41593-021-00940-5}{10.1038/s41593-021-00940-5} \\

GOD
& fMRI, behavioral
& Healthy adults
& Open
& \href{https://doi.org/10.1038/ncomms15037}{10.1038/ncomms15037} \\

ZuCo
& EEG, eye-tracking
& Healthy adults
& Open
& \href{https://doi.org/10.1093/gigascience/giz082}{10.1093/gigascience/giz082} \\

\hline
\end{tabular}
}
\end{sidewaystable}
 
\bibliographystyle{plainnat}

\bibliography{refs}

\end{document}